\documentclass[aps,showpacs,preprintnumbers,amsmath,amssymb, 
twocolumn, tightenlines,
]{revtex4}

\usepackage[dvips]{graphicx}\usepackage{bm}
\usepackage{amsmath}
\usepackage{dcolumn}
\usepackage{bm}

\begin{document}

\bibliographystyle{apsrev} 

\title {On bound states of multiple t-quarks due to Higgs exchange }

\author{M.Yu.Kuchiev} \email[Email:]{kmy@phys.unsw.edu.au}
\author{V.V.Flambaum} \email[Email:]{flambaum@phys.unsw.edu.au}

\affiliation{School of Physics, University of New South Wales, Sydney
  2052, Australia}

\author{E. Shuryak} \email[Email:]{shuryak@tonic.physics.sunysb.edu}

\affiliation{Department of Physics, State University of New York,
 Stony Brook, NY 11794, USA}

    \date{\today}

    \begin{abstract} 
Froggatt, Nielsen et al \cite{Froggatt} suggested that the Higgs boson exchange between top quarks produces enough attraction to generate their multiple bound states. Furthermore they
 claimed that the system of 6 top and 6 anti-top quarks
is bound so strongly that the binding energy 
nearly compensates the masses of t-quarks, making it very light.
We calculated  the  energy of such states more accurately, variationally and by self-consistent mean field approximation, and found that these state are weakly bound
for massless Higgs boson and unbound with the account for its mass.

    \end{abstract}
    

    \maketitle


Froggatt, Nielsen et al (see all refs   \cite{Froggatt}) suggested
interesting possibility that the Standard Model
Higgs boson coupling to top quarks is large enough to
generate a whole  spectroscopy of new bound states. Furthermore,
they thought that for large enough number
of quarks this binding can be strong enough to
stabilize top quarks, as bound neutrons are stabilized
in nuclei.

The particular state discussed in \cite{Froggatt} is for the
number of quarks $N=12$, 6 t-quarks and 6 anti-t-quarks,
with all spin and color values,
 which all occupy the same lowest 1s orbital. 
In a simple approximation 
used in \cite{Froggatt} the binding energy for this state
is so large that the total mass of this T-ball turns to be zero,
inside claimed accuracy.
 The authors of \cite{Froggatt} made proposals of how to find 
such states at Tevatron and LHC. One can even consider dark matter made from T-balls.


In the Standard Model the interaction of t-quarks with the Higgs boson
 is proportional to the large mass of the t-quark,
 $m_{\mathrm{t}}=172.6$ GeV, with a coupling $g_{\,\mathrm{t}}\simeq
 0.989 \pm 0.008$ \cite{Froggatt}.  Therefore the effective
Coulomb coupling is about as strong as $\alpha_s$ at such
scale, with the additional advantage that quarks and antiquarks
of any color are equally attracted by the Higgs exchange.
Thus binding should grow with the number of quarks.

 The calculations
  of \cite{Froggatt} provided only a rough estimate for the
 binding energy, which was
based on analogy with Bohr energies of the Hydrogen atom.
However, an atom has an attractive Coulomb center, while for
multi-top bound states considered the charge is smeared out over the volume of 
the multi-top. The more accurate calculations of 
the present work show that this makes huge difference 
for 12 t-quarks, which is not deeply bound at all, but rather unbound,
if realistic limits on the Higgs mass is used.

We use first the variational approach, which is simple and analytic,
and then a self-consistent mean-field approach.
 The many-body and recoil corrections are expected to be small, $\sim
 1/N$, where $N=12$ is the number of t-quarks, making accuracy of our
 binding energy $\sim10$\%. For weak coupling the
 non-relativistic approximation  is certainly valid.
 The interaction between t-quarks due to Higgs boson exchange may be described by the following Hamiltonian ($\hbar=c=1$):
 \begin{equation}\label{H}
H= \sum_{i=1}^N \frac{p_i^2}{ 2m_{\mathrm{t}} } +
 \sum_{1\le i< k}^N V(|{\bf r}_i-{\bf r}_k|)
\end{equation}
where $V(r)$ is 
\begin{equation}
V(r) \,=\, - \frac{\alpha_\mathrm{H} }{r}\,\exp{(-m_\mathrm{H} \,r)}~.
	\label{V}
\end{equation}
Here $m_\mathrm{H}$ is the Higgs mass, and the coefficient $\alpha_\mathrm{H}$ is introduced to account for the strength of the Higgs boson exchange. The
t-channel exchange in the t-quark scattering, 
 following \cite{Froggatt}, leads to the effective Coulomb
coupling $\alpha_\mathrm{H}=g_{\,\mathrm{t}}^2/(8\pi)$. It is stated in \cite{Froggatt} that the inclusion of
s- and u- channels makes the effective interaction between t-quarks twice as strong 
\cite{foot}. In line with this assessment  let us presume that
\begin{equation}
	\alpha_\mathrm{H}\,\simeq \,g_{\,\mathrm{t}}^2/(4\pi)\,\approx 1/(4\pi)~,
	\label{al}
\end{equation}
where the mentioned value $g_{\,\mathrm{t}}\simeq 0.989$ was employed.

 The variational approach assumes that the wave function of the
 multi-top system incorporates a product of the $N$ orbitals $\psi(r)$ for $N$ t-quarks. In order to describe our idea it suffices to take $\psi(r) \propto \exp{(-q \,r)}$, where $q$ is a variational parameter (self-consistent calculations, which allow one to find $\psi(r)$ more accurately, are mentioned below). Then from Eq.(\ref{H}) we find the variational boundary for the multi-top energy
\begin{equation}
\label{E}
\langle H \rangle =N \frac{q^2}{2m_{\mathrm{t}}}- \frac{5}{16} N(N-1) \,\alpha_\mathrm{H} \,q\,f(y)~,
\end{equation}
where $f(y)=({1+4y/5 +y^2/5})/{(1+y)^4}$, $y=m_\mathrm{H}/(2q)$.
 The estimates of  \cite{Froggatt} neglect Higgs mass, so
 $m_\mathrm{H}=0$. Due to this reason,
 let us also start with this case, when we have $y=0,f(0)=1$.
 The minimum of  $\langle H\rangle $ is achieved for $q=(5/16)(N-1) m_{\mathrm{t}} \alpha_\mathrm{H}$ reading
\begin{equation}\label{EC}
\langle H \rangle \,=\,- k\,N(N-1)^2\,\alpha_\mathrm{H}^2\, m_{\mathrm{t}}~.
\end{equation}     
The found binding energy  grows with $N$ as  $\sim N^3$,
 in line with the main idea of \cite{Froggatt}. However, we find that the coefficient
in Eq.(\ref{EC})  $k = 25/512 \simeq 0.049$
 is $\simeq 5$ times smaller than the coefficient 1/4 of
 the Hydrogen-atom model used in \cite{Froggatt}. 
Qualitatively, a quark inside is influenced by an effective attractive charge that is smaller than 
the charge in the Hydrogen-type model.

 We now proceed to the mean field approach describing it by
 the self-consistent Hartree-Fock equations, which determine more accurately the probing single-particle wave function $\psi(r)$. Using this approach one finds from Eq.(\ref{H})
\begin{equation}
	\langle H \rangle \,=\,N K+\frac{1}{2}\,N(N-1)\,U~,
	\label{HHF}
\end{equation}
where $K$ and $U$ describe the kinetic and potential energies
\begin{align}
&K\,=\,\frac{1}{2m_\mathrm{t}}\int |\,\nabla\psi(r)\,|^2\,d^3r~,
\label{K}
\\
&U\,\,=\int V(|{\bf r}-{\bf r}'|)\,\psi^2(r)\,\psi^2(r')\,d^3r\,d^3r'~.
\label{U}
\end{align}
By variation of the function $\psi(r)$ in Eq.(\ref{HHF}) one obtains the self-consistent equation on the single-particle wave function $\psi(r)$ and the corresponding single-particle energy $E$
\begin{align}
& E\psi(r)=\left(-\frac{\Delta}{2m_\mathrm{t}}+(N-1)W(r)\right)\psi(r)~,
\label{HFPsi}
\\
&W(r)\,=\int V(|{\bf r}-{\bf r}'|)\,\psi^2(r')\,d^3r'~.
	\label{HFW}
\end{align}
Solving Eqs.(\ref{HFPsi}),(\ref{HFW}) one then recovers the total energy of the system from Eqs.(\ref{HHF}),(\ref{K}) and (\ref{U}). 

\begin{figure}[tbh]
  \centering \includegraphics[ height=5cm, keepaspectratio=true]{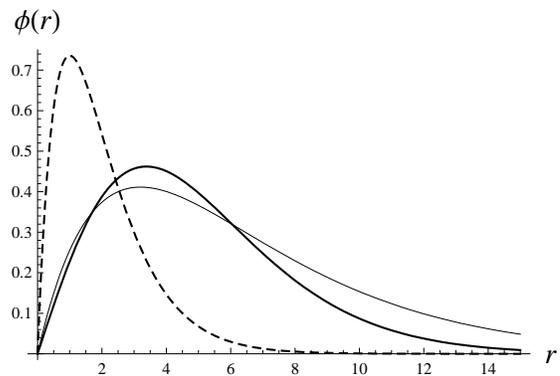}
\vspace{0cm}
\caption{ 
The radial single-particle wave function $\phi(r)=r \,\psi(r)$ versus the distance $r$ from the multi-top center for $m_\mathrm{H}=0$.
Thick line - solution of the self-consistent Hartree-Fock Eqs.(\ref{HFPsi}),(\ref{HFW}), thin line - a variational solution $\phi(r)\propto r \exp(-qr))$ used in Eq.(\ref{E}),
broken line - 1s function in the Hydrogen-like model, 
all in units of $a_0=\hbar/[\,(N-1)m_\mathrm{t}c \alpha_\mathrm{H}]$.
}\label{one} 
   \end{figure}

The results of our calculations for the case of $m_\mathrm{H}=0$ are shown in Fig.1, which presents the radial functions 
$\phi(r) = r \,\psi(r)$ with normalization $\int_0 ^\infty \phi^2(r)\,dr=1$. 
The figure clearly illustrates the fact that wave functions derived in this work deviate strongly from the Hydrogen-type model, compare the two solid lines with the broken one that represents the Hydrogen model.
The distance in Fig. \ref{one} is measured in  units of $a_0=1/[\,(N-1)m_\mathrm{t} \alpha_\mathrm{H}]$, which represents an effective Bohr radius in the problem. 
For $N=12$ and $\alpha_\mathrm{t}$ from Eq.(\ref{al}) this effective radius is close to the Compton radius of the t-quark,  $a_0\simeq 1.14/m_\mathrm{t}$. 
\noindent

These calculations confirm that the total energy of the multi-top is
indeed described by Eq.(\ref{EC}), in which the coefficient $k$ is
small, $k=0.05426$ (within $10$ \% agrees
 with the previously obtained variational
value $k=25/512\simeq 0.049$). Note that $k$ is a number, which does
not depend on any parameters of the problem: so 
 regardless of the value of $\alpha_\mathrm{H}$ in Eq.(\ref{al})
or other details,
we conclude that the binding energy of the multi-top is  smaller,
 by a factor of 4.6, than the binding energy in the Hydrogen-type
 model.  The larger radius of the wave
 function shown in Fig.\ref{one} provides another illustration for this statement. 
The maximum of the wave function that comes out of the self-consistent calculation is located at $r_\mathrm{max}=3.38 a_0$, which for $N=12$ equals $r_\mathrm{max}=3.86/m_\mathrm{t}$.
Thus, even neglecting the Higgs mass, the binding energy of the multi-top represents only $\sim 4$\% of the sum of $N$ t-quark  masses.  The size of the multi-top is substantially larger than the Compton radius of the t-quark.


\begin{figure}[b]
  \centering \includegraphics[ height=5cm, keepaspectratio=true]{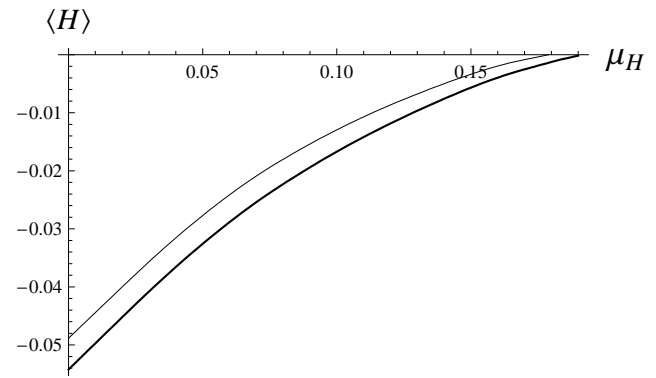}
\vspace{0cm}
\caption{ 
The total energy $\langle H\rangle$ of the multi-top versus the mass of the Higgs boson $m_\mathrm{H}$, energy is measured in units of $E_0=N(N-1)^2\,\alpha_\mathrm{H}^2\, m_{\mathrm{t}}c^2$ 
(compare Eq.(\ref{EC})), mass of the Higgs boson is represented by $\mu_\mathrm{H} = m_\mathrm{H} c a_0/\hbar$.
Thick line - solution of the Hartree-Fock equations, thin line - a variational solution $\psi(r)\propto \exp(-qr)$.}\label{two} 
   \end{figure}
\noindent

The large size of such bound state makes it necessary to take into
account that the interaction in Eq.(\ref{H}) has a finite range
$1/m_\mathrm{H}$.  We found -- both from Eq.(\ref{E}) and from the
Hartree-Fock Eqs.(\ref{HFPsi}),(\ref{HFW}) --
that the bound state of the multi-top disappears if $m_\mathrm{H}$  is sufficiently large.
Figure \ref{two} illustrates the rise of the total energy with
increase of the Higgs boson mass (in the Coulomb units used in
Fig.\ref{two} the dependence on the number of the t-quarks does not
manifests itself). The found from this data critical value of the
Higgs mass, which makes the bound state impossible, proves small
compared to existing experimental lower limits. For $N=12$ t-quarks this state disappears when $m_\mathrm{H}\ge 0.157 \,m_\mathrm{t}\simeq 27$ Gev, if the variational Eq.(\ref{E})
is used. Calculations based on the mean field approach of Eqs.(\ref{HFPsi}),(\ref{HFW}) give a close result $m_\mathrm{H}\ge 0.168 \,m_\mathrm{t}\simeq 29$ Gev. 

We conclude that  to make the multi-top with $N=12$ t-quarks bound
 one needs $m_\mathrm{H}\le 29$ GeV, which
contradicts  available experimental lower limits
 for the Higgs boson mass. This means that the bound state of the
 multi-top cannot exist for $ 2\le N\le 12$ t-quarks due to the
 short-range character and insufficient strength of the interaction. 
Inclusion of strong force (gluons) will make bound states possible
even for one $\bar t t$ pair, but with very small binding.

We end up with a discussion of some pertinent questions.
First,  the strong  $N^3$ dependence of energy Eq. (\ref{EC})
on the number of particles $N$ leaves room for a speculations
of what would happen at larger $N>12$, in particular for
$N=24$ tops, closing 2s state.

Second, suppose there exist new generations of heavy fermions which would allow one
to consider larger $N$, or for the matter of argument
masses of, say, $b$ quarks are changed to that of the top.
Another possibility is to consider a ball containing  $W$ and $Z$ bosons.
In this case the number of bosons $N$ may be arbitrary large, therefore, a
simple non-relativistic model  based on Eq. (\ref{EC}) would predict a
spontaneous creation of additional $W^+$, $W^-$ and $Z$ bosons and collapse
of the ball. One may also consider a mixed fermion-boson ball. 

Simple estimates show that the total binding energy $\langle H\rangle $ in Eq. (\ref{EC}) becomes comparable with the total rest
mass for $N>50$. 
A correct solution of the problems mentioned requires self-consistent relativistic treatment for quarks and bosons, including the nonlinearity of the Higgs potential:  we hope to
report related results shortly.

One of us (ES) thanks Holger Nielson for presentation of his
ideas, some of which this paper addresses.
The work of VF and MK is supported by the Australian Research Council
and that of ES by US-DOE grants DE-FG02-88ER40388 and
DE-FG03-97ER4014.


\begin{thebibliography}{99}
    
\bibitem{Froggatt} C.D. Froggatt,
 L.V. Laperashvili, R.B. Nevzorov, H.B. Nielsen, C.R. Das, arxiv: 0804.4506
[hep-ph]; 
C.D. Froggatt,  H.B. Nielsen,   L.V. Laperashvili,arxiv : hep-ph/0406110,
 Int.J.Mod.Phys A20, 1268 (2005);
C.D. Froggatt, H.B. Nielsen,
arxiv : hep-ph/0308144, Surveys High Energy Phys. 18,55 (2003).

\bibitem{foot}
One can raise questions related to this strong enhancement, asking in particular whether 
an enhancing  factor in the effective interaction is close to 2, as it follows from the 2008 work in \cite{Froggatt} (or is it slightly less than that, as was alleged previously), 
but we put them aside, since our results show that even with this large factor 2 the Higgs 
boson exchange is not able to justify the binding in the multi-top. 

 \end{thebibliography}
\end{document}